\documentclass[a4paper,fleqn,usenatbib]{mnras}

\usepackage{graphicx}   
\usepackage{amsmath}    
\usepackage{amssymb}    
\usepackage{subfig}
\usepackage{color}

\usepackage[T1]{fontenc}
\usepackage{ae,aecompl}

\title[Tidal tails of Galactic GCs]{The eye of Gaia on globular clusters structure: tidal tails}
\author[Sollima et al.]{A. Sollima$^{1}$\thanks{E-mail:
antonio.sollima@inaf.it}\\
$^{1}$ INAF Osservatorio di Astrofisica e Scienza dello spazio di Bologna, 
via Gobetti 93/3, 40129 Bologna, Italy\\
}

\date{Accepted 2020 April 25. Received 2020 April 22; in original form 2020 January 23}

\pubyear{2020}

\begin{document}
\label{firstpage}
\pagerange{\pageref{firstpage}--\pageref{lastpage}}
\maketitle


\begin{abstract}
I analyse the projected density distribution of member stars over a wide area 
surrounding 18 Galactic globular clusters using the photometric and astrometric 
information provided by the second data release of the Gaia mission. 
A 5D mixture modelling technique has been employed to optimally isolate the signal
of the cluster stellar population from the contamination of the Galactic field, taking advantage 
of its different distribution in the space formed by colours, magnitudes, parallaxes and 
proper motions.
In 7 clusters I detect collimated overdensities at a $>3~\sigma$ level 
above the background density extending well beyond the cluster tidal radius, 
consistent with the distortion expected as a result of the tidal interaction 
with the Milky Way potential.
In five of these clusters (NGC288, NGC2298, NGC5139, NGC6341 and NGC7099) 
spectacular tidal tails extend up to the border of the analysed field of view 
at 5 degrees from the centre.
At large distances from the cluster centre, the orientation of the detected 
overdensities appears to be systematically aligned with the cluster orbital path, 
in agreement with the predictions of N-body simulations.
The fraction of stars contained in the tidal tails of these clusters is also used to determine the first observational 
estimate of their present-day destruction rates.
\end{abstract}


\section{Introduction}
\label{intro_sec}

Any stellar system immersed in a gravitational potential is subject to its 
tidal strain. In particular, the presence of the external potential introduces 
a repulsive centrifugal term which adds to the internal gravitational potential 
and determines an anisotropic shape of the effective potential contours.
The effect of tidal distortion is more pronounced at large distances from the 
center where tidal forces dominates over the internal gravitation.
Globally, the zero-velocity surfaces have a characteristic elongated shape 
(Roche lobe) with two saddle points (the so-called "Lagrangian points", 
located at a distance from the cluster centre called "Jacobi radius") and aligned 
with the direction of the galactic potential gradient 
\citep[i.e. toward the galactic centre;][]{binney87}.
Stellar orbits confined in the outer region of the stellar system are therefore 
distorted toward the Lagrangian points and the stellar density 
follows the shape of the Roche lobe. 
A star with orbital energy exceeding the effective potential at the Largangian 
points can escape from the system if its orbit crosses an aperture around 
these points, whose size depends on its energy excess \citep{fukushige00}.
Once escaped, the star initially moves away from the satellite following the 
path along its escape direction i.e. radially along the satellite-galactic 
centre direction. As the distance of the star increases, the 
Coriolis force becomes significant pushing the star toward the same orbital 
path of its original stellar system.
The appearance of the whole system is therefore characterised by the presence 
of a pair of symmetric tails emerging from the satellite and following a S-shape 
\citep{montuori07,klimentowski09}.
The stars in the tidal tails become independent satellites with 
integrals of motion similar to those of their original host stellar system, 
and therefore follow its orbit with a slightly different orbital period. 
Stars escaped at early epochs accumulate a large phase delay and are therefore 
more distant from the satellite than those recently escaped.
So, the orientation and prominence of tidal tails provide crucial information on 
the orbit and the recent mass-loss history of a stellar system.

Examples of tidal features have been observed in interacting galaxies 
\citep{duc15} and around the most recent accretion events occurred in the Milky Way and M31 
\citep{ibata94,ibata01}. All these systems are characterised by large mass-loss 
rates \citep[up to $10^{8}~stars~Gyr^{-1}$;][]{fardal06,law10} leaving observational 
features easily detectable with imaging techniques as surface 
brightness or star counts excess 
\citep{belokurov06a,martinez10}.

The detection of tidal tails is more challenging in star 
clusters, characterised by a significantly smaller mass-loss rate 
\citep[$\sim 10^{2\div 4}~stars~Gyr^{-1}$;][]{dinescu99}.
The stars in the tails constitute indeed a tiny fraction ($<0.1\%$) of those 
contained in the region surrounding the cluster, mainly populated by Galactic 
interlopers. 
For this reason, filtering techniques are used to maximize the signal 
of the cluster by assigning weights to stars according to their position in 
the colour-magnitude diagram \citep[the "matched-filter";][]{rockosi02}.
Thanks to this techniques, it has been possible to detect tidal features around 
many globular \citep[GCs;][]{grillmair95,leon00,chun10,chen10,jordi10,chun15,shipp18,carballo18} 
and open clusters \citep{bergond01,dalessandro15}.
Most of these studies discovered only deformations of the density contours close 
to the clusters' tidal radii interpreted as the result of tidal effects.
Spectacular exceptions are constituted by the $30^{\circ}$-long tail detected around 
Palomar 5 \citep{odenkirchen01,erkal17} and, to a less extent, the tidal tails detected 
in NGC 5466 \citep{belokurov06b}, Palomar 1, \citep{niederste10}, Palomar 14 \citep{sollima11}, 
Eridanus and Palomar 15 \citep{myeong17} and NGC 7492 \citep{navarrete17}.

The 2nd data release of the Gaia mission \citep{gaia18a} recently provided (beside colours and magnitudes) 
parallaxes and proper motions for $\sim1.3\times 10^{9}$ stars across the entire sky, 
allowing to increase the number of dimensions of the parameter space where to 
select cluster members.
Thanks to this dataset it has been possible to detect extended tidal tails in 
NGC5139 \citep{ibata19a}, NGC288 \citep{kaderali19}, NGC362 \citep{carballo19}, NGC3201 \citep{bianchini19}, NGC5904 
\citep{grillmair19} and around a few open clusters \citep{roser19a,roser19b,tang19}.

In this paper I perform a systematic search for tidal tails around a sample of 18 
nearby GCs using the full 5D parameter space provided by Gaia data.
In Sect. \ref{data_sec} the analysed sample of GCs, together with the description 
of the adopted dataset is presented. 
The algorithm adopted to determine the projected density map is described in Sect. \ref{alg_sec} and 
the results of the analysis are presented in Sect. \ref{res_sec}. 
Finally, the conclusions are presented in Sect. \ref{concl_sec}.

\section{Observational material}
\label{data_sec}

\subsection{GC sample}
\label{sample_sec}

\begin{table}
 \centering
  \caption{GCs analysed in this work (column (1)). The detection of 
  significant tidal tails, according to the criteria described in Sect. \ref{map_sec} is reported in column (2) and the destruction rate, 
  where available, is listed in column (3).}
  \begin{tabular}{@{}lcc@{}}
  \hline
 NGC & tidal tails & $\nu$\\
     &             & $Gyr^{-1}$\\
 \hline
288  & Y & 0.033$\pm$0.003\\
1851 & N & \\
1904 & N & \\
2298 & Y & 0.057$\pm$0.007\\
2808 & N & \\
3201 & N & \\
4590 & N & \\
5139 & Y & 0.019$\pm$0.002\\
5272 & N & \\
5897 & N & \\
5904 & Y & \\
6205 & N & \\
6341 & Y & 0.018$\pm$0.001\\
6362 & Y & \\
6752 & N & \\
7078 & N & \\
7089 & N & \\
7099 & Y & 0.085$\pm$0.004\\
\hline
\end{tabular}
 \label{table1_tab}
\end{table}

The GCs analysed in this paper have been selected as those with the largest 
probability of detection of low-surface brightness features.
Many factors affect this probability, among them: {\it i)} the amount of mass 
lost in recent epochs, {\it ii)} the fraction of the cluster stellar population sampled by the adopted catalog, 
{\it iii)} the density of Galactic field stars, and {\it iv)} the separation of cluster and field 
stars in the considered parameter space.

Although theoretical predictions of the mass-loss rate of GCs exist in the 
literature \citep{gnedin97,allen06}, many uncertainties 
affect these estimates (due to the uncertainties in the adopted Galactic potential, 
GC space velocities, recipes for mass loss, etc.). 
So no selection has been made on the basis of this parameter.
On the other hand, even in case of large mass loss rates, the tidal tails of 
low-mass and/or distant GCs will be sampled only with a small number of stars.
For this reason, I excluded GCs with integrated apparent magnitude $V>9.5$. 

The fraction of sampled cluster stellar population depends mainly on the 
fraction of Main Sequence stars brighter than the limiting magnitude of 
Gaia and therefore on the cluster heliocentric distance.
The GCs analysed here were selected among those at distances $d_{\odot}<15$ 
kpc \citep[from][2010 edition]{harris96}.
Considering the Gaia limiting magnitude ($20<G_{cut}<21$; see Sect. 
\ref{gaia_sec}), this corresponds to $\sim$ 1 magnitude below the turn-off 
of a typical old and metal-poor ($t>10 ~Gyr$; $[Fe/H]<-1$) 
GC stellar population \citep{bressan12}. 

To limit the fraction of Galactic disc and bulge interlopers, I excluded those GCs 
lying at latitudes $|b|<10^{\circ}$ and at a projected distance from the 
Galactic centre $R_{GC}<30^{\circ}$. 
I also excluded NGC104 and NGC362 because of the contamination from Small 
Magellanic Cloud stars in the background of these clusters. 

Some nearby GCs, in spite of their projected position in the sky, have a mean 
proper motions significantly different from that of the surrounding Galactic 
population. The stellar population of these GCs can be easily isolated allowing 
an efficient analysis.
Among these GCs I included in the sample NGC3201 (with a systemic proper motion 
different by $\Delta\mu\sim14~mas~yr^{-1}$ from the mean proper motion of the 
field population), in spite of its low latitude ($b=8^{\circ}.64$).

Instead, the GC NGC5286 has been excluded because of the extremely high small-scale 
spatial variation of the Gaia catalog completeness even at bright magnitudes 
($G_{cut}<20$; see Sect. \ref{gaia_sec}). 

The final sample consists of 18 GCs (see Table \ref{table1_tab}), $\sim$12\% of the 
entire Milky Way GC system, and constitutes one of the most extensive sample adpoted 
for this kind of studies.

\subsection{Gaia catalog selection}
\label{gaia_sec}

The analysis performed here is entirely based on data provided by the Gaia 2nd data 
release \citep{gaia18a}.
This survey measured parallaxes, proper motions, magnitudes and colours for 
$1.3\times 10^{9}$ stars in the entire sky, including almost all known Galactic 
GCs in both hemispheres.
The catalog has a formal limiting magnitude of $G=21$, although this value depends 
on celestial position.
Moreover, the source completeness varies with a patchy structure characterised by 
source density fluctuations that reflect the scan law pattern of the survey \citep{arenou18}.

For each GC, the full set of information of all sources contained in a circle with 
radius of $5^{\circ}$ centered on the GC centre has been retrieved.
It consists of $G$ magnitudes, $G_{BP}-G_{RP}$ colours, parallaxes ($p$), proper 
motions ($\mu_{\alpha}^{*},~\mu_{\delta}$) together with their associated 
uncertainties and corresponding covariances.
Uncertainties in parallaxes have been corrected using the prescriptions of \citet{lindegren18}.
The $G$ magnitudes and $G_{BP}-G_{RP}$ colours have been corrected for interstellar extinction using the 
reddening maps of \citet{lallement19}\footnote{Although the reddening maps of \citet{lallement19} are 
calculated using only stars up to 3 kpc from the Sun, at Galactic latitudes 
$|b|>10^{\circ}$ they include more than 99.9\% of the dust column density along 
the line of sight of all the GCs analysed in this work.} and the extinction coefficients by 
\citet{casagrande18}.

To maximize the detection efficiency avoiding to introduce any bias, no selection 
has been made on the basis of the Gaia quality flags.
It has indeed been shown that the 2nd Gaia data release is 
characterized by spatial variation of the astrometric accuracy occurring at 
both small and large scales (from a few arcminutes to several degrees) with a 
patchy distribution across the sky, 
depending on the local stellar density and the satellite scanning law 
\citep{arenou18}. So, any cut made on the basis of astrometric quality 
parameters would spuriously decrease the stellar density of the less surveyed 
regions. On the other hand, the inclusion of stars with poorly measured 
astrometric/photometric parameters increases the noise in the density 
determination, so that the cost-effectiveness of the application of quality cuts 
is not trivial to be assessed. As a test, I compared the density maps of the GC 
NGC 5139, where tidal tails were already detected using the same dataset 
adopted here \citep{ibata19a}, derived with and without the application of 
various cuts in parallax/proper motion errors: the inclusion of quality cuts, 
while requiring a fine tuning to avoid the emergence of patches parallel to the 
Gaia scanning law, produces no significant improvement in the detection of tidal tails.
So, the only selection applied to the catalog has been made on the basis of the dereddedned
$G$ magnitude. Indeed, for the reliability of the present analysis, it is crucial to ensure 
a constant source detection efficiency across the analysed field of view, and 
the completeness fluctuations of Gaia at its faint end can significantly 
affect the results of the analysis.
The adopted magnitude cut ($G_{cut}$) varies from cluster to cluster and has 
been set to ensure a smooth variation of the density across the analysed field of view.
For this purpose, the following procedure has been adopted:
\begin{itemize}
\item{For different guesses of $G_{cut}$, the sources with $|G-G_{cut}|<0.25$ have been selected;}
\item{The field of view has been divided in $6\arcmin\times6\arcmin$ bins 
evenly distributed within an annulus with internal and external radii of 
$1^{\circ}$ and $5^{\circ}$ around the cluster centre, and the number of selected sources contained in each bin has been counted;}
\item{The logarithm of star counts has been fitted by a first order polynomial in (X, Y, $\log N$) 
and the r.m.s of the fit has been calculated;}
\item{The largest value of $G_{cut}$ ensuring a $r.m.s.<0.05~dex$ has been adopted.}
\end{itemize}
The values of $G_{cut}$ determined according to the above procedure lie in the range 
$20<G_{cut}<21$, with brighter cuts in those GCs located in densely populated 
regions (i.e. close to the Galactic plane and/or the bulge) or characterised by a small 
number of Gaia passages.

Corrected distances ($X,~Y$) and proper motions have been calculated 
using an orthographic projection of the canonical celestial coordinates and proper motions 
\citep[see eq. 2 of][]{gaia18b}.

\section{Method}
\label{alg_sec}

\begin{figure*}
 \includegraphics[width=\textwidth]{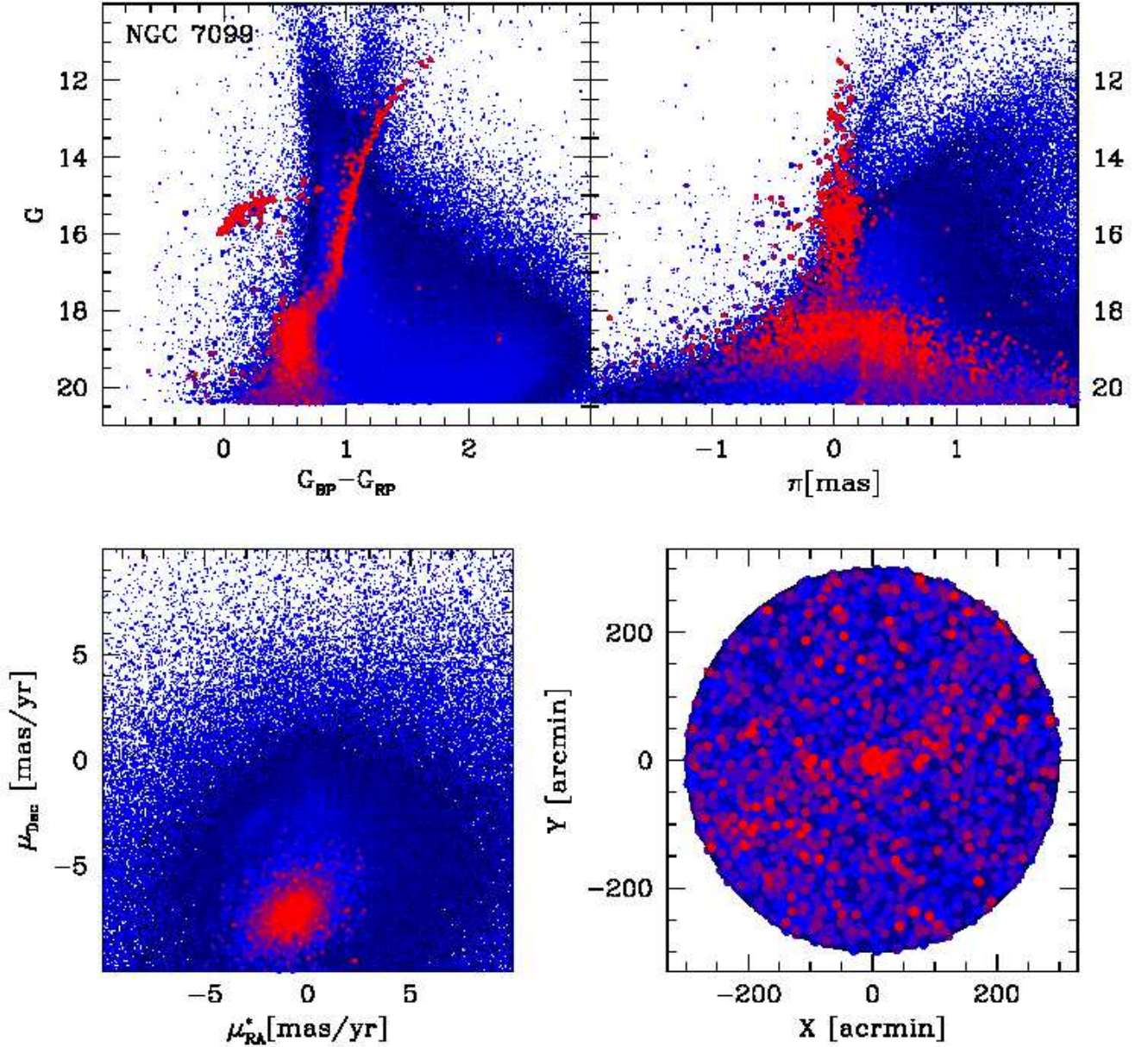}
 \caption{Distribution of NGC 7099 stars in several planes of the parameter space.
 Top-left: $G_{BP}-G_{RP},G$ colour-magnitude diagram; top-right: parallax-$G$ magnitude 
 plane; Bottom-left: proper motions plane; Bottom-right: projected positions map.
 Stars with increasing cluster-to-field probability contrast $\Delta \ln P$ are plotted with 
 colours from blue to red (grey to black in the printed version of the paper).}
\label{sel}
\end{figure*}

The algorithm adopted in this work is a 5D mixture modelling \citep{mclachlan88}.
Schematically, the distribution of a representative sample of cluster members 
and Galactic field stars in the parameter space is modelled with suitable analytic 
functions. The relative normalization of the model cluster and field distributions 
is calculated through a maximum-likelihood technique 
using samples of stars located in different region of the analysed 
field and used to determine the local density of cluster stars.

The stars contained within (outside) $\Delta R<0.5^{\circ}$ from the cluster centre 
\citep[from][]{goldsbury13}, $\Delta \mu<2~mas~yr^{-1}$ from the systemic cluster motion 
\citep[from][]{baumgardt19}\footnote{for NGC5139, because of its large extent and 
velocity dispersion, these conditions have been set to $\Delta R<0.8^{\circ}$ and 
$\Delta \mu<2.3~mas~yr^{-1}$} and $\Delta p<5\epsilon_{p}$ from the measured cluster 
parallax \citep[from][; where $\epsilon_{p}$ is the parallax error]{gaia18b} 
have been selected as {\it reference samples} for the cluster (field) population.

The parallax distribution of cluster stars has been assumed to be a Dirac delta centered on the 
mean cluster parallax. This is a reasonable assumption since the parallax differences 
among cluster stars are more than 100 times smaller than the typical parallax uncertainty 
in all the GCs of the sample.
So, the probability ($P_{p,i}^{c}$) for a star to be a cluster member, on the basis of its parallax only, 
is given by
\begin{equation}
\ln P_{p,i}^{c}=-\frac{(p_{i}-\langle p\rangle_{c})^{2}}{2~\epsilon_{p,i}^{2}}-
\ln(\epsilon_{p,i})
\label{pc_eq}
\end{equation}
where $p_{i}$ and $\epsilon_{p,i}$ are the parallax of the i-th star and its associated 
uncertainty and $\langle p\rangle_{c}$ is the mean cluster parallax in units of $mas$.
The distribution of field star parallaxes has been fitted with the empirical function
\begin{equation}
\Gamma(p)=\left(1+\frac{p}{a_{1}}\right)^{-\alpha_{1}}\left(1+\frac{p}{a_{2}}\right)^{-\alpha_{2}}
\label{par_eq}
\end{equation} 
The best fit values of $a_{1}$, $a_{2}$, $\alpha_{1}$ and $\alpha_{2}$ have been determined 
using a Monte Carlo scheme.
For any guess of the $a_{1}$, $a_{2}$, $\alpha_{1}$ and $\alpha_{1}$ parameters,
$N_{f}$ (equal to the number of {\it reference sample} field stars) synthetic 
parallaxes have been randomly extracted from the corresponding $\Gamma(p)$ 
distribution (eq. \ref{par_eq}).
Real and synthetic star parallaxes have been sorted  
and a gaussian shift with dispersion equal to the parallax uncertainty of each 
star has been added to the corresponding particle.
The cumulative distributions of real and synthetic parallaxes have been then 
compared using a Kolmogorov-Smirnov test
and the values of the parameters providing the largest KS probability have been chosen.
The probability for a star to be a field star, on the basis of its parallax, 
is given by
\begin{multline}
\ln P_{p,i}^{f}=\ln\int_{0}^{+\infty}\Gamma(p)\exp\left[\frac{-(p-p_{i})^{2}}{2\epsilon_{p,i}^{2}}\right]~dp-\\
\ln\left(\sqrt{2\pi}\epsilon_{p,i}\int_{0}^{+\infty} \Gamma(p)~dp\right)
\label{pf_eq}
\end{multline}

The distribution of cluster stars in the proper motions space has been 
modelled with a 2D Gaussian with a decreasing dispersion as a function of distance from the cluster centre
according to the prediction of the \citet{king66} model best fit of \citet{deboer19}.
The normalization of the model proper motion dispersion profile has been chosen by maximising the
log-likelihood
\begin{eqnarray}
\ln L&=&\sum_{i=1}^{N} \ln P_{\mu,i}\nonumber\\
\ln P_{\mu,i}&=&-\frac{1}{2}[\delta X_{i}^{2}+\delta Y_{i}^{2}-2\tilde{\rho_{i}}\delta X_{i}\delta Y_{i}+\ln(1-\tilde{\rho_{i}}^{2})+\nonumber\\
&&\ln(s_{\mu X,i}^{2}s_{\mu Y,i}^{2})]-\ln(2\pi)\nonumber\\
\label{like_eq}
\end{eqnarray}
where
\begin{eqnarray}
\delta X_{i}^{2}&=&\frac{(\mu_{\alpha,i}^{*}-\langle\mu_{\alpha}^{*}\rangle_{c})^{2}}{(1-\tilde{\rho_{i}}^{2})s_{\mu X,i}^{2}}\nonumber\\
\delta Y_{i}^{2}&=&\frac{(\mu_{\delta,i}-\langle\mu_{\delta}\rangle_{c})^{2}}{(1-\tilde{\rho_{i}}^{2})s_{\mu Y,i}^{2}}\nonumber\\
s_{\mu X,i}^{2}&=&\epsilon_{\mu \alpha,i}^{2}+\sigma^{2}(R_{i})\nonumber\\
s_{\mu Y,i}^{2}&=&\epsilon_{\mu \delta,i}^{2}+\sigma^{2}(R_{i})\nonumber\\
\tilde{\rho_{i}}&=&\frac{\rho_{i}~\epsilon_{\mu \alpha,i}\epsilon_{\mu \delta,i}}{s_{\mu X,i}s_{\mu Y,i}}\nonumber\\
\label{muc_eq}
\end{eqnarray}
among the $N=N_{c}$ {\it reference sample} cluster stars. 
In the above equations, 
$\langle\mu_{\alpha}^{*}\rangle_{c}$ and $\langle\mu_{\delta}\rangle_{c}$ are the systemic 
cluster proper motions, $\mu_{\alpha,i}^{*}$ and $\mu_{\delta,i}$ are the proper motions of 
the i-th star, $\epsilon_{\mu \alpha,i}$, 
$\epsilon_{\mu \delta,i}$ are the proper motion errors, $\rho_{i}$ is the 
correlation coefficient of the two uncertainties and $\sigma(R_{i})$ is the 
normalised model proper motion dispersion at the projected distance of the i-th 
star (in units of $mas~yr^{-1}$).
For the field population, the proper motion distribution has been modelled with a 2D 
tilted gaussian function with parallax-dependent dispersion
\begin{equation}
\sigma(p)=\frac{a}{p}~\{1+\exp[b(\log p-c)]\}
\label{pm_eq}
\end{equation}
The values of the free parameters $a$, $b$, $c$ and $\theta$ have been derived 
by maximising 
the log-likelihood of eq. \ref{like_eq} applied to the {\it reference sample} field stars 
($N=N_{f}$) and using 
\begin{eqnarray}
\delta X_{i}^{2}&=&\frac{[(\mu_{\alpha,i}^{*}-\langle\mu_{\alpha}^{*}\rangle_{f}) \cos\theta+(\mu_{\delta,i}-\langle\mu_{\delta}\rangle_{f}) \sin\theta]^{2}}{(1-\tilde{\rho_{i}}^{2})(s_{X,i}^{2}+\sigma^{2}(p_{i}))}\nonumber\\
\delta Y_{i}^{2}&=&\frac{[-(\mu_{\alpha,i}^{*}-\langle\mu_{\alpha}^{*}\rangle_{f}) \sin\theta+(\mu_{\delta,i}-\langle\mu_{\delta}\rangle_{f}) \cos\theta]^{2}}{(1-\tilde{\rho_{i}}^{2})(s_{Y,i}^{2}+\sigma^{2}(p_{i}))}\nonumber\\
s_{X,i}^{2}&=&\epsilon_{\mu \alpha,i}^{2} \cos^{2}\theta+\epsilon_{\mu \delta,i}^{2} \sin^{2}\theta+\rho_{i} \sin{2\theta}\epsilon_{\mu \alpha,i}\epsilon_{\mu \delta,i} \nonumber\\
s_{Y,i}^{2}&=&\epsilon_{\mu \alpha,i}^{2} \sin^{2}\theta+\epsilon_{\mu \delta,i}^{2} \cos^{2}\theta-\rho_{i} \sin{2\theta}\epsilon_{\mu \alpha,i}\epsilon_{\mu \delta,i} \nonumber\\
\tilde{\rho_{i}}&=&\frac{(s_{Y,i}^{2}-s_{X,i}^{2}) \sin{2\theta}+ 2(\rho_{i}s_{X,i}s_{Y,i} \cos{2\theta})}{2\sqrt{(s_{X,i}^{2}+\sigma^{2}(p_{i}))(s_{Y,i}^{2}+\sigma^{2}(p_{i}))}}\nonumber\\
\label{muf_eq}
\end{eqnarray}
where $\langle\mu_{\alpha}^{*}\rangle_{f}$ and $\langle\mu_{\delta}\rangle_{f}$ are the mean systemic proper 
motion of {\it reference} field stars in the two directions \footnote{Note that the simplified functional forms of the parallax 
(eq. \ref{par_eq}) and proper motion (eq. \ref{pm_eq}) distribution of field 
stars were chosen to adequately fit the observed distribution of {\it reference} 
field stars without adding a large number of free parameters. This choice is justified by the fact that, 
while a proper modelling of these distributions would require to account for the 
projection of the density and velocity ellipsoids of the various Galactic 
components along the various line of sights, as well as the Gaia selection function, the small-scale 
variations of such distributions are smoothed by the relatively large observational errors.}.

The density of {\it reference sample} 
cluster stars at the position of the i-th star in the $(G_{BP}-G_{RP}),G$ colour-magnitude diagram ($\Lambda_{CMD}^{c}$)
has been calculated using a k-nearest neighbour algorithm (with k=10) and assuming a 
metric between colours and magnitudes 
$(\Delta_{G_{BP}-G_{RP}},\Delta_{G})=(1,6)$ i.e. corresponding to the average 
ratio of standard deviations in colours and magnitudes of {\it reference sample} cluster stars 
in the considered sample of GCs.
\begin{equation*}
\Lambda_{CMD_i}=\frac{10}{\pi \left[\Delta_{G_{BP}-G_{RP}}^{2}+\left(\frac{\Delta_{G}}{6}\right)^{2}\right]}
\end{equation*}
where $\Delta_{G_{BP}-G_{RP}}$ and $\Delta_{G}$ are the colour and magnitude difference between the i-th star and its 10-th nearest neighbour.
The corresponding cluster membership probability is given by 
\begin{equation}
\ln P_{CMD,i}^{c}=\ln\Lambda_{CMD,i}^{c}-\ln N_{c}
\label{cmdc_eq}
\end{equation}
The field probability has been calculated using same algorithm applied to the 
{\it reference sample} of field stars with $\Delta~\log p<0.1$ from the 
considered star 
\begin{equation}
\ln P_{CMD,i}^{f}=\ln\Lambda_{CMD,i}^{f}-\ln N_{f,i}
\label{cmdf_eq}
\end{equation}
where $N_{f,i}$ is the number of {\it reference sample} field stars contained in the 
interval encompassing the parallax of the i-th star
\footnote{It should be noted that, at odds with the parallax and proper motion probabilities 
calculated convolving 
intrinsic distributions with individual uncertainties, the probability linked to the
position in the CMDs includes the global effect of 
photometric uncertainties neglecting individual differences (i.e. I assigned the same 
probability to all stars lying in the same portion of the CMD, regardless of their 
individual colour/magnitude uncertainties). Again, this approximation is made to 
avoid a complex modelling of the intrinsic distribution in the 
colour-magnitude-parallax space which depends on many uncertain parameters 
(like star formation history, mass function, photometric completeness, mass-segregation, etc.).}.

The total membership probability to the cluster (field) population has been obtained 
from eq.s \ref{pc_eq}, \ref{like_eq}, \ref{muc_eq} and \ref{cmdc_eq} 
(and \ref{pf_eq}, \ref{like_eq}, \ref{muf_eq}, \ref{cmdf_eq}, respectively) as
\begin{eqnarray*}
\ln P_{i}^{c}&=&\ln P_{p,i}^{c}+\ln P_{\mu,i}^{c}+\ln P_{CMD,i}^{c}\nonumber\\
\ln P_{i}^{f}&=&\ln P_{p,i}^{f}+\ln P_{\mu,i}^{f}+\ln P_{CMD,i}^{f}\nonumber\\
\end{eqnarray*}
Stars with a larger difference $\Delta \ln P\equiv \ln P^{c}-\ln P^{f}$ are those 
lying in a region of the parameter space with a larger contrast between cluster 
and field distributions (see Fig. \ref{sel}).

The probabilities of stars located in different regions of the analysed field of 
view have been then used to derive the relative fraction of cluster-to-field stars.
For this purpose, a grid with knots separated by $6\arcmin$ in both directions has 
been defined and stars contained within a projected distance $R_{lim}$ from each 
knot have been selected. In two cases (NGC5272 and NGC5904) a nearby GC is 
present within the field of view (NGC5466 and Pal 5, respectively). For these cases, 
the area surrounding the nearby clusters has been masked (see Fig. \ref{map1} and \ref{map2}).
In each subsample of stars, the cluster-to-field ratio ($\eta$) has been chosen as the one 
maximizing the likelihood
\begin{equation*}
\ln L=\sum_{i=1}^{N_{xy}} \eta~\ln P_{i}^{c}+(1-\eta) \ln P_{i}^{f}
\end{equation*}
and the corresponding density of cluster members is given by
\begin{equation}
\Sigma(X,Y)=\frac{\eta N_{xy}}{A(R_{lim})}
\label{dens_eq}
\end{equation}
where $N_{xy}$ and $A(R_{lim})$ are the number of stars in the subsample and its 
corresponding area.

The choice of $R_{lim}$ should be adaptive to ensure a better resolution in regions with 
a large number of cluster members and a large smoothing in poorly populated regions. 
I adopted in each knot of the grid the minimum value of $R_{lim}$ providing 
$\eta~N_{xy}>100$. 

The above procedure should in principle provide a background-subtracted 
density map. On the other hand, an implicit assumption is that the adopted distribution of 
field stars in the parameter space does not vary within the field of view.
While this is a reasonable assumption in many GCs of the sample, in GCs located 
in regions characterised by steep variation of the field population properties 
(i.e. those at low 
Galactic latitudes and/or close to the bulge) this approximation leads to a gradient 
in the derived density. However, given the relatively small size of the considered 
field of view, such a gradient has a small amplitude and varies smoothly across the field of view.
For this purpose, the logarithm of the 2D density of regions at distances $R>2^{\circ}$ from the cluster 
centre has been fitted with a first-order polynomial in the $(X,~Y,~\log \Sigma)$ space 
and subtracted.

The significance of the derived density has been estimated using a Monte Carlo 
technique. A sample of $10^{3}$ extractions has been made by randomly 
reshuffling the position angles of all sources and calculating the corresponding 
density map. The r.m.s. of the various density determinations at distances 
$R>2^{\circ}$ from the cluster centre has been adopted as the typical uncertainty.

\section{Results}
\label{res_sec}

\subsection{Density maps}
\label{map_sec}

\begin{figure*}
 \includegraphics[width=\textwidth]{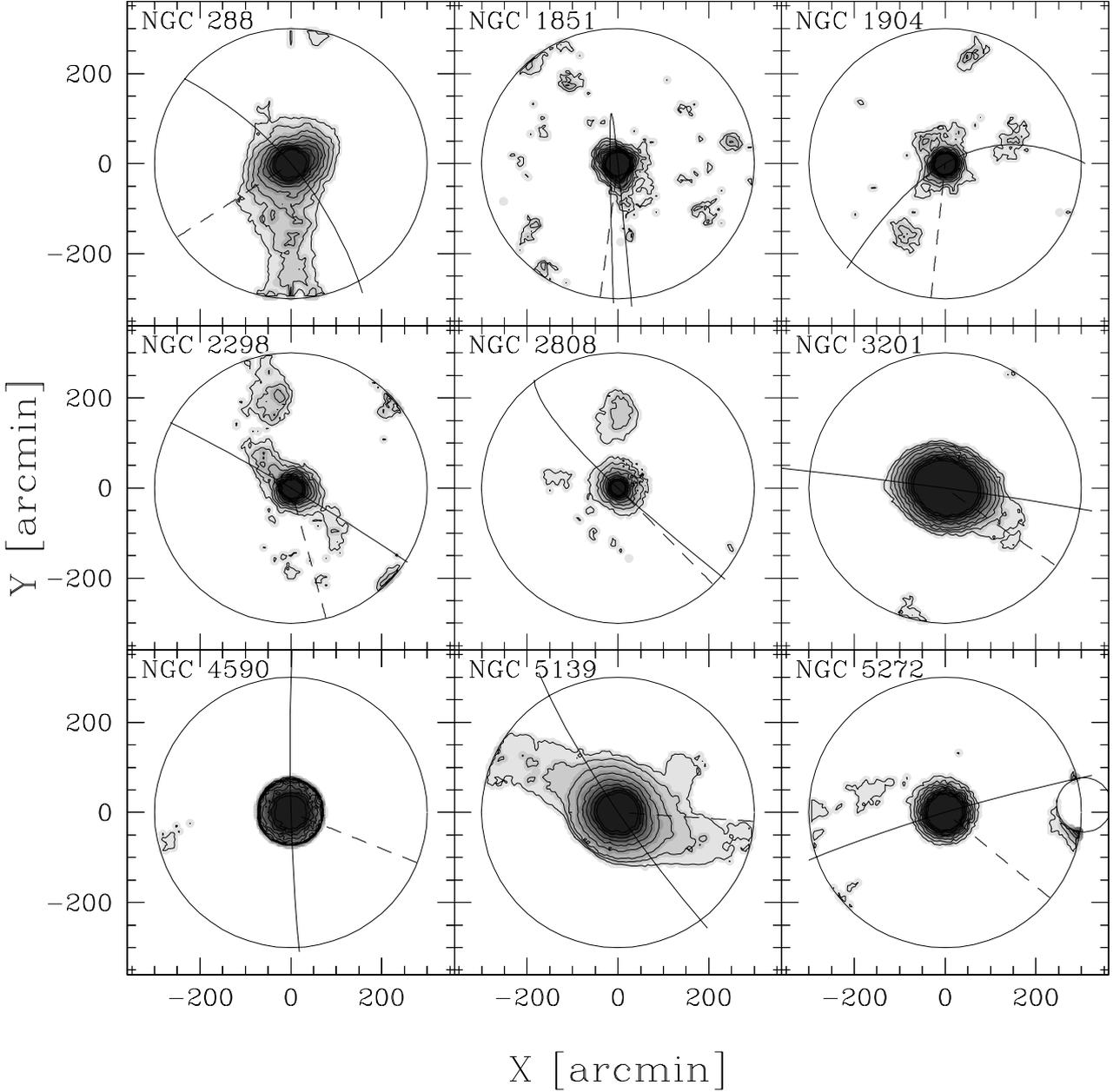}
 \caption{Projected density map for the GCs NGC288, NGC1851, NGC1904, NGC2298, NGC2808,
  NGC3201, NGC4590, NGC5139 and NGC5272. Contours from 2 to 10$\sigma$ above the 
  background density are plotted in logarithmic scale in steps of 1$\sigma$. The cluster orbital path 
  and the direction to the Galactic centre are overplotted as solid and dashed 
  lines, respectively.}
\label{map1}
\end{figure*}

\begin{figure*}
 \includegraphics[width=\textwidth]{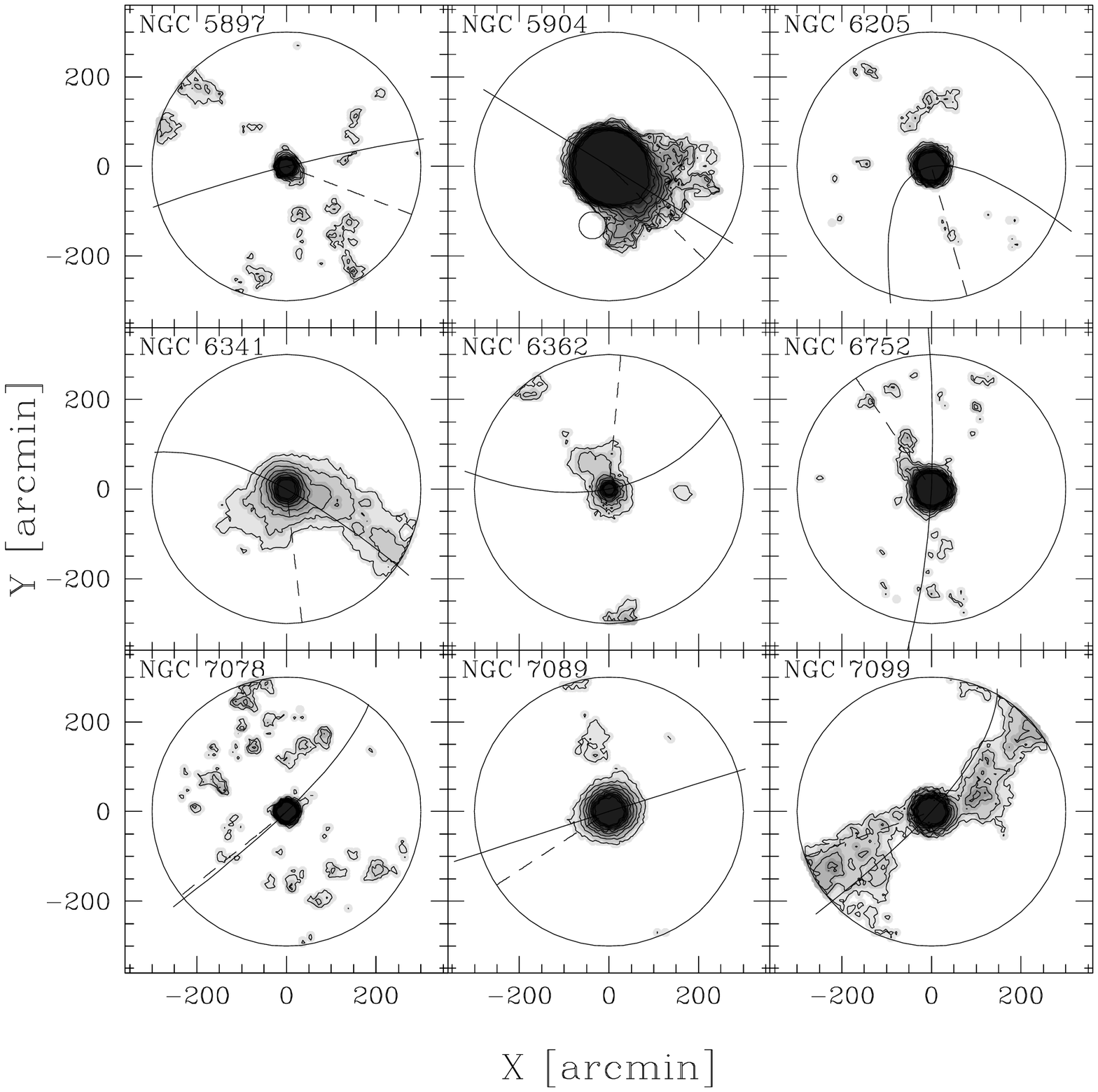}
 \caption{Same as Fig. \ref{map1} for NGC5897, NGC5904, NGC6205, NGC6341, NGC6362, NGC6752,
  NGC7078, NGC7089 and NGC7099.}
\label{map2}
\end{figure*}

The projected density maps for the 18 GCs of the sample are shown in Fig.s \ref{map1} and 
\ref{map2}.
It is apparent that, while the density contours near the cluster centres are extremely 
regular, at large distances many clusters show anisotropic density distributions.

None of the previous works on this subject defined a rigorous criterion to claim the detection 
of tidal tails.
Following the definition of tidal tails as "a collimated overdensity of stars emerging 
from the cluster in opposite directions" \citep{toomre72} I defined the two following criteria:
\begin{itemize}
\item{The closed iso-density contour at 3$\sigma$ above the background density containing the 
cluster centre must exceed a distance of 1.5 times the orbit-averaged Jacobi radius;}
\item{The integrated projected density measured in opposite sectors must be significantly 
larger than that measured in the complementary region of the sky outside the Jacobi radius.}
\end{itemize}

The first criterion implies the presence of the cluster population outside the Roche lobe.
The orbit-averaged Jacobi radius ($R_{J}$) has been calculated by averaging 
over the orbit the Jacobi radius estimated using eq. A2 of \citet{allen06} and 
the GC masses by \citet{baumgardt19}. The cluster orbit has been integrated 
within the Galactic potential of \citet{johnston95} using a fourth-order 
Runge-Kutta integrator adopting the present-day systemic velocities listed by 
\citet{baumgardt19}.

The second criterion requires that the observed overdensity is preferentially oriented in 
one direction. For this purpose the annulus between $R_{J}$ and $2~R_{J}$ has been divided in two 
complementary regions constituted by opposite sectors of $90^{\circ}$ width centered at a 
given position angle \citep[see Fig. 5 of][]{sollima11}. For each position angle, the ratio 
between the densities (measured using eq. \ref{dens_eq}) within each region has been 
calculated, and the position angle providing the maximum ratio ($\mathcal{R}_{max}$) has been chosen. 
The same estimate has been made in the Monte Carlo randomized
extractions (see Sect. \ref{alg_sec}) and compared with $\mathcal{R}_{max}$.
The alignment is considered statistically significant if $\mathcal{R}_{max}$ is found to be larger 
than the same value measured in at least 997 (out 1000) Monte Carlo extractions. 

Seven GCs pass the above criteria: NGC288, NGC2298, NGC5139, NGC5904, NGC6341, NGC6362 
and NGC7099. Other GCs (like NGC3201 and NGC6752) show hint of tidal 
deformation at a low ($\sim~2\sigma$) significance level.
Note that, among the 7 GCs with positive detection, in 5 cases (NGC288, NGC2298, 
NGC5139, NGC6341 and NGC7099) the tidal tails extend up to the border of the analysed field of view 
(at $5^{\circ}$ from the cluster centre) and in two cases (NGC5139 and NGC7099) they remain 
significant at both sides of the cluster covering an extent of $\sim10^{\circ}$.

It is also interesting to note that 4 (out 7) GCs show asymmetric tidal tails. 
Such an effect is expected in GCs close to their apo/peri-centres: in this situation, 
the velocity of stars varies quickly with the orbital phase so that stars in the 
leading (trailing) tail move away from the cluster while those in the trailing (leading) tail slowly 
approach the cluster when it leaves (reaches) the extreme of the orbit. However, 
while all the 4 GCs exhibiting asymmetric tails are relatively close to their 
apo/peri-centers, in only 2 of them the observed elongation is directed toward the 
expected direction. 
So, a more likely explanation is that this
evidence is due to statistical fluctuations occurring close to 
the threshold density which make appear only one side of the tails.

\subsection{Comparison with literature works}

The GCs analysed in this work have been already studied in the past using different photometric 
datasets. 
In particular, the tidal tails of NGC 288 have been discovered up to the same distance 
from the cluster centre and with the same orientation found in the present 
study by \citet{shipp18} using photometric 
data from the Dark Energy Survey, and \citet{kaderali19} using Gaia data.
Previous studies \citep{grillmair95,leon00,piatti18} also found low-significance distortion close 
to the tidal radius of this cluster. 

\citet{ibata19a} used Gaia data to detect the same tidal feature 
observed here around NGC5139, which appears to be extended for $\sim28^{\circ}$ 
and connected with the previously discovered Fimbulthul stream \citep{ibata19b}.
The detection of tidal tails around NGC5139 was also claimed by \citet{leon00} but questioned by 
\citet{law03} because of the effect of differential reddening neglected by the former study.

Similarly, \citet{grillmair19} found a $\sim50^{\circ}$-extended tidal tails around NGC5904 using
Gaia data selected over and area $>6000$ sq.deg.
Also in this case, small tidal distortions in NGC5904 were previously reported by \citet{leon00} and \citet{jordi10}.

Some density excess close to the tidal radii of NGC2298 \citep{leon00,balbinot11,carballo18} 
and NGC7099 \citep{chun10} have been reported in the past. However, none of them reach a 
comparable level of significance and extent with respect to that found in the present 
work.

No previous studies have found any extra-tidal feature around NGC6362 and NGC6341. 

Among the other GCs analysed here, low-amplitude overdensities outside the tidal radii have been claimed in
NGC1851 \citep{leon00,carballo18,shipp18}, NGC1904 
\citep{grillmair95,leon00,carballo18,shipp18}, NGC2808 
\citep{grillmair95,chen10,carballo18}, NGC3201 
\citep{grillmair95,chen10,bianchini19}, NGC4590 \citep{grillmair95}, NGC5272 \citep{leon00}, NGC6205 
\citep{leon00}, NGC7008 \citep{grillmair95,jordi10,chun10} and NGC7089 \citep{grillmair95}.
It is not easy to compare the results presented in this paper with those of the 
above studies, since all of them use different thresholds and smoothing levels and in most cases 
the identification of extra-tidal features is made on the basis of a qualitative inspection of the density map. 
Consider also that the lack of detection in the present study of significant tidal tails around 
these GCs could be due to the limiting magnitude of Gaia ($20<G_{cut}<21$; see Sect. \ref{gaia_sec}), significantly brighter than 
those of many of the above studies. This is particularly important in distant GCs where, in spite 
of the lack of astrometric information, their deep 
photometry allows to sample the cluster stellar population better by a factor of 10 or 
more, even in the intrinsically poorly populated periphery of GCs. 

\subsection{Tidal tails orientation}

\begin{figure*}
 \includegraphics[width=\textwidth]{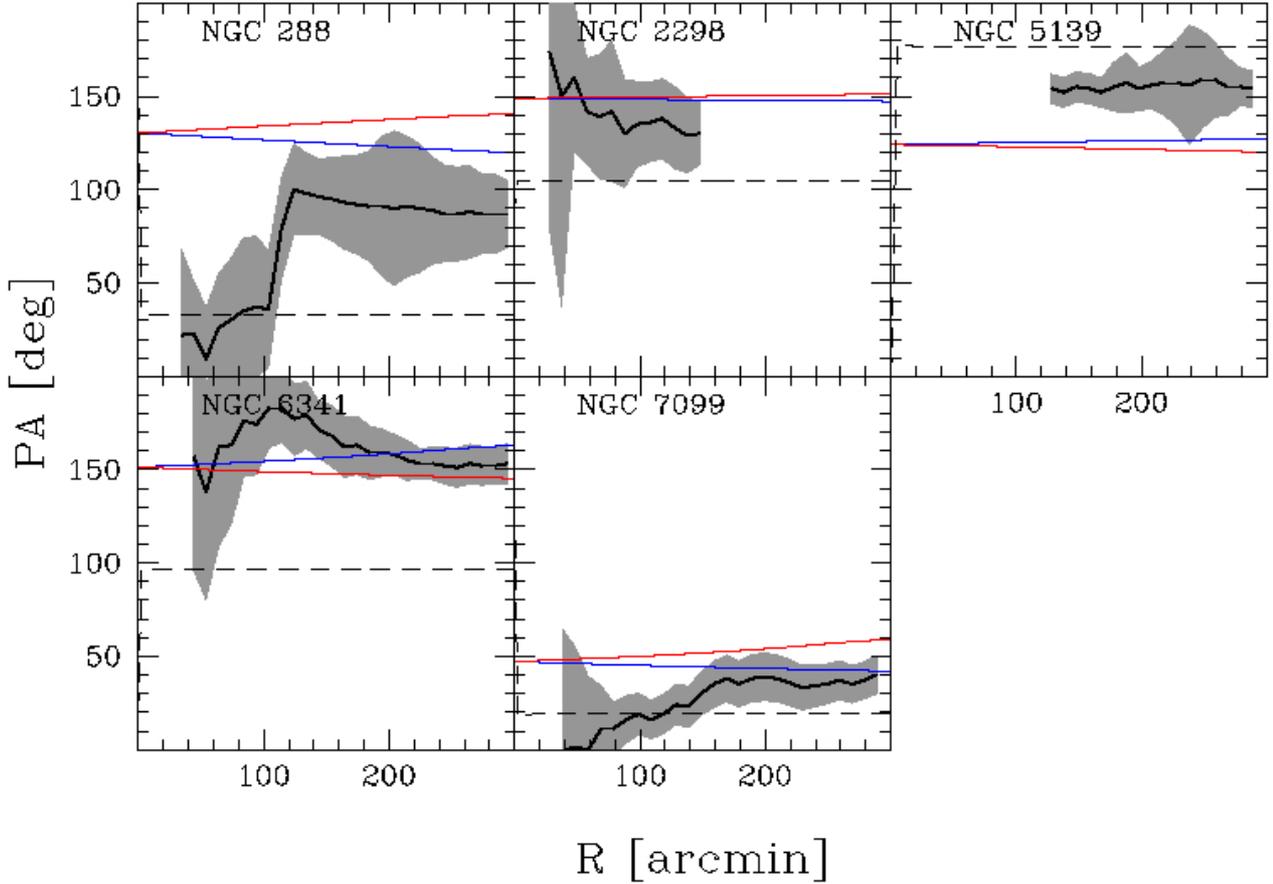}
 \caption{Position angle of the tidal tails as a function of the projected distance from the 
 cluster centre (black solid lines) for NGC288, NGC2298, NGC5139, NGC6341 and NGC7099. 
 The shaded area mark the 1$\sigma$ uncertainty.
 The direction of the leading (blue lines; dotted lines in the printed version of the paper) 
 and trailing (red lines; dot-dashed lines in the printed version of the paper) sides of the orbit 
 and that to the Galactic centre (black dashed lines) are overplotted.}
\label{ang}
\end{figure*}

As introduced in Sect. \ref{intro_sec}, because of the different directions of the 
forces acting on stars within/close to the Jacobi radius and at 
large distance from the cluster centre, tidal tails are expected to show a characteristic 
S-shape, with the innermost part elongated toward the Galactic centre and a
torsion toward the cluster orbital path at large distances \citep{montuori07,klimentowski09}.

Qualitatively, this prediction seems to be verified in many of the GCs showing extended
overdensities outside their tidal radii (see Fig.s \ref{map1} and \ref{map2}).
In particular, the inner density contours of NGC288, NGC5139, NGC5904, 
NGC6362 and NGC7099 (and with a smaller significance also in NGC3201 and NGC6752) 
appear to be elongated in the direction of the Galactic centre. On the other hand, in 
NGC288, NGC6341 and NGC7099, where it is possible to detect the 
tidal tails up to several degrees, the orientation of the overdensity tends to rotate toward 
the cluster orbital path.

To quantitatively test the above hypothesis, in the 5 GCs showing the most extended extra-tidal 
overdensities (NGC288, NGC2298, NGC5139, NGC6341 and NGC7099), the field of view outside 
$0.5~R_{J}$ has been divided in annuli of $0.1^{\circ}$ width and the background-subtracted 
density of each annulus as a function of the position angle has been fitted with a cosine function.
The angle corresponding to the phase of maximum of the best fit curve is plotted as a 
function of the distance from the centre of each of the 5 considered GCs in Fig. 
\ref{ang}.
In all cases, the observed orientation is located in the portion of the plane contained 
between the direction of the orbit and that of the Galactic centre.
In NGC288 and NGC7099 a drift of the tidal tails position angle from the direction of 
the Galactic centre to that of the cluster orbit at increasing distances is noticeable.
 
A $\chi^{2}$ test applied to regions at $R>R_{J}$ in each individual GC 
indicates probabilities 
$66\%<P_{\chi^{2}}<86\%$ that the distribution of position angles follows the 
same trend of the orbit\footnote{For this test, the averages between the leading 
and trailing directions of the orbit have been used.}. On the other hand, if all the 
5 GCs are considered together, the same test gives a probability of 99.8\%, indicating a 
significant general alignment of the outer portion of the tails with the orbital paths in this 
sample of GCs.

\subsection{Destruction rates}

\begin{figure*}
 \includegraphics[width=\textwidth]{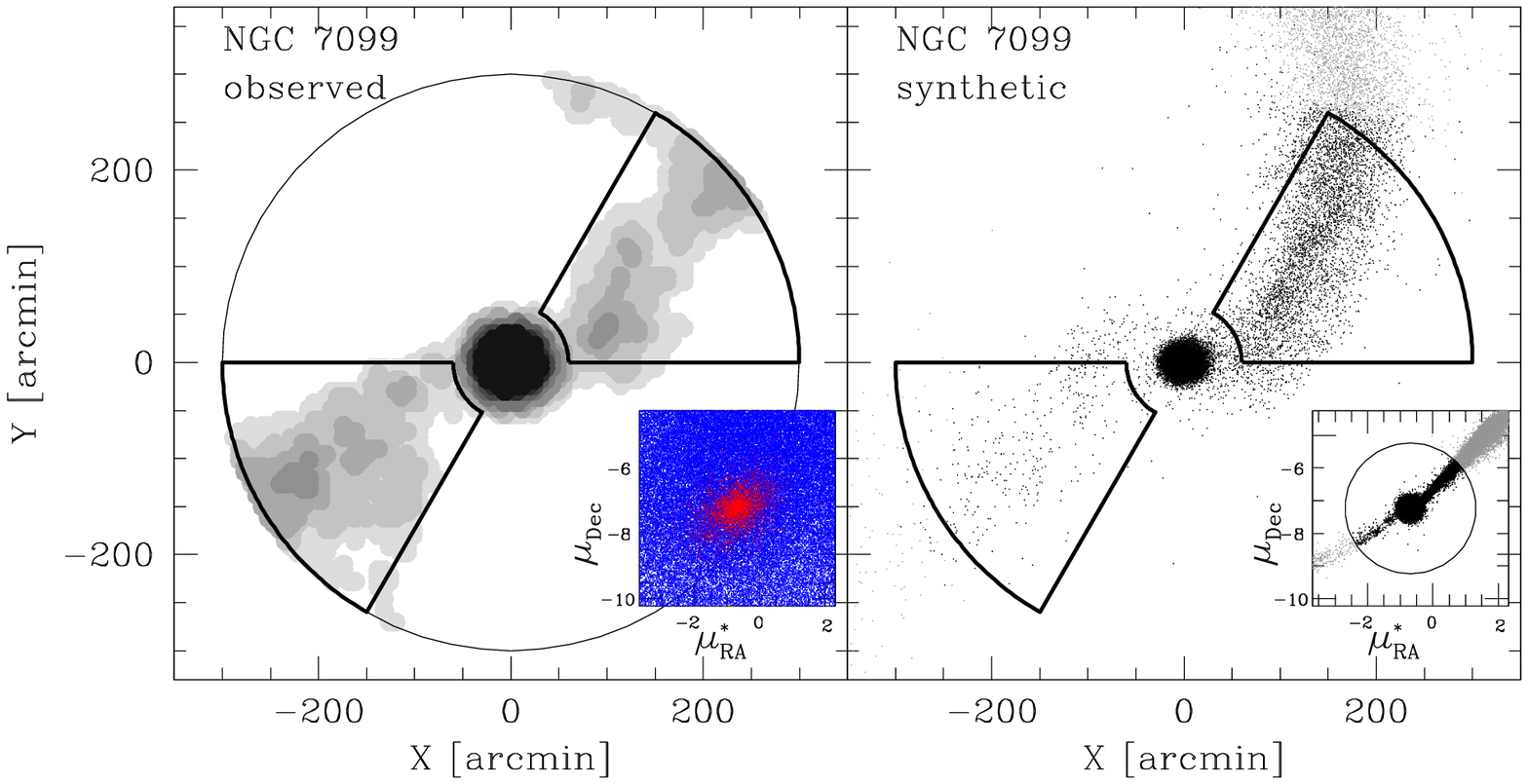}
 \caption{Left panel: observed density map of NGC7099. The proper motion probability contrast 
 map is plotted in the inset using the same colour code of Fig. \ref{sel}. 
 Right panel: projected distribution of synthetic particles of the corresponding simulation. 
The proper motion distribution is shown in the inset. Black and grey dots 
correspond to particles within and outside the proper motion selection box. The adopted selection 
boxes are overplotted to both panels as thick black lines.
 }
\label{streak}
\end{figure*}

The stars contained in the tidal tails have been lost in recent epochs by the GCs.
Although, after its expulsion, the distance of a star from the cluster centre 
follows a non-monotonic variation with time according to the eccentricity of 
the cluster orbit \citep{kupper10,kupper12}, the phase delay of groups 
of stars evaporated at different epochs increases with time, so that stars in 
the tails at increasing distance from the centre were on average expelled at 
earlier epochs.
It is therefore possible to link the fraction of stars contained in the tidal tails with 
the rate at which a cluster looses its mass through 
a comparison with suitable models.

For this purpose, as a first step I calculated the number of 
stars contained in the tidal tails ($N_{obs}^{tails}$) and within the cluster 
tidal radius ($N_{obs}^{in}$) in the 5 GCs with the most prominent tidal 
tails (NGC288, NGC2298, NGC5139, NGC6341 and NGC7099).
The first quantity has been calculated by integrating the background-subtracted 
projected density within two $60^{\circ}$-wide regions located at the opposite sides of each 
cluster at distances $R>1.5~R_{J}$ encompassing the tidal tails (see Fig. \ref{streak}).
Unfortunately, the same approach cannot be adopted in the cluster central region 
characterised by high stellar density and consequently low completeness.
So, the number of stars inside the cluster has been estimated by normalizing the 
projected density profile of the 
\citet{king66} model best fit of \citet{deboer19} to the density measured 
in the outer portion of the cluster, which should not be affected by 
significant crowding. The total number of stars has been then calculated by 
integrating the model profile from the center to $R_{J}$.

The mass-loss history of each GC has been modelled using the 
technique of streaklines \citep{kupper12}. Briefly, the orbit of each cluster has 
been reconstructed and, at each integration step, a 
particle with a small ($\Delta E<0.1~km^{2} s^{-2}$) positive energy has been put 
at one of the Lagrangian points and with a velocity directed outward from the cluster.
The orbit of each ejected particle and that of the GC (simulated as a point mass 
particle) have been then followed together within the Galactic potential for 2 Gyr 
till the presend-day position of the cluster. I neglected the cluster mass 
variation since stars in the portion of the tails analysed here are those 
ejected at recent epochs in a short time interval (from 0.12 to 1 Gyr depending on 
the cluster orbit), during which the clusters lost a small ($<6\%$ in all cases) 
of their present-day mass.
This approach is faster than a 
canonical N-body simulation (since the orbit of the escaped stars are followed 
individually neglecting mutual interactions and with the relatively long 
timestep required to accurately integrate their orbit), and has been proven to 
be effective in reproducing the structure and evolution of tidal tails 
\citep{kupper12,mastrobuono12}.
As already assumed in Sect. \ref{map_sec}, the cluster orbits have been 
integrated within the \citet{johnston95} Galactic potential, using the cluster 
masses and present-day positions and velocities of \citet{baumgardt19}.
The (constant) mass-loss rate of each simulation has been set to $10^{4}~stars~Gyr^{-1}$, 
which ensures to simulate a statistically significant number of tidal tails stars in a 
reasonable computation time.
At the end of the simulation, the positions and velocities of simulated 
particles have been converted into equatorial coordinates, proper motions and 
radial velocities. The particles with proper motions within 
$\Delta\mu<2~mas~yr^{-1}$ from the systemic cluster motion and contained within 
the same regions defined above for observations, have been counted providing 
the number of predicted stars in tidal tails ($N_{sim}^{tails}$) for the assumed 
mass-loss rate.

The fractions of mass lost per unit time (the so-called "destruction rate") have 
been therefore calculated as
\begin{equation*}
\nu=\frac{10^{4} N_{obs}^{tail}}{N_{obs}^{in} N_{sim}^{tails}}
\end{equation*}
and are listed in Table \ref{table1_tab}. 

Of course, many approximations are made in the above estimate: real clusters 
lose stars non-uniformly during their evolution, and the evolution of the cluster 
and Galactic potentials are neglected.
More importantly, two-body relaxation lead to a preferential loss of low-mass 
stars, so that the mass function of in the tails is expected to be significantly different from that 
within the cluster tidal radius \citep{balbinot18}. So, the estimated mass-loss rates are lower limits to the actual values.
In this context, two-body relaxation proceeds in each cluster on different 
timescales, so that this effect can affect different clusters in different ways. 
However, the mass range sampled by Gaia is relatively small 
($0.6<M/M_{\odot}<0.8$) and similar in all the considered GCs, so this effect should affect the above estimate only at a second order.

The derived destruction rates range between $0.018<\nu/Gyr^{-1}<0.085$, corresponding to 
dissolution times between 12 and 55 Gyr. 
The destruction rates derived here agree with the 
predictions calculated for the same GCs of \citet{gnedin97} (using Fokker-Planck simulations) and direct N-body simulations (H. Baumgardt; 
private communication), while they are significantly 
larger than those reported by \citet{allen06}.
It should be considered that all these theoretical works adopt different Galactic potentials, cluster orbits and filling factors, 
so a detailed comparison is not possible.

According to a large number of studies \citep{gnedin97,balbinot18} based on 
theoretical arguments and simulations, the mass loss rate of a star cluster is determined 
by the combination of internal (due to two-body relaxation) and external (due to 
tidal shocks experienced during passages across the Galactic disk and at the peri-Galacticon) 
effects. The two above processes have different dependences on the cluster characteristics: 
the internal process is expected to produce an almost constant mass loss rate 
every half-mass relaxation time \citep{spitzer87}, while the external process depends on the 
orbital characteristics \citep{ostriker72,aguilar88}.
Unfortunately, because of the small number of GCs with significant tidal tails, it 
is not possible to perform a meaningful analysis of correlations with the parameters 
affecting the mass loss. 
However, at face value, the Pearson correlation coefficients 
between the logarithm of the destruction rate and those of the 
half-mass relaxation time \citep[from][]{baumgardt19} and 
tidal shocks-related destruction rate 
\citep[given by the sum of eq.s 1 and 2 of][]{dinescu99} turn out 
to be -0.60 and 0.36, respectively.
While the small sample size makes both correlations not significant, 
they are expected 
since in GCs with short relaxation times two-body relaxation has efficiently 
pushed the energy of their stars beyond the escape threshold, and GCs with 
strong tidal interactions are easily destroyed.
 
\section{Conclusions}
\label{concl_sec}

I reported the results of a comprehensive analysis of the projected density 
map of a sample of 18 Galactic GCs using the set of astrometric and photometric data
provided by the Gaia 2nd data release.
The use of all the available information in the 5D space composed by parallaxes, 
proper motions and magnitudes allowed to optimally separate the signal of 
the cluster population from the contamination from fore/background Galactic field 
interlopers. This allowed to sample the cluster population down to very faint 
surface brightness levels.

The derived maps show significant deviation from sphericity close to the tidal radii 
of many analysed GCs. For 7 (out 18) GCs these overdensities are statistically 
significant up to distances exceeding the size of the Roche lobe being likely produced 
by the strain exherted by the Galactic tidal forces.
In 5 GCs the detected overdensities extend up to several degrees from the cluster centre 
(extending over $10^{\circ}$ in NGC5139 and NGC7099) and show the typical shape of symmetric tidal tails emerging 
from the cluster.
While for a few GCs (NGC288, NGC5139, NGC5904) these features were already 
reported by previous studies \citep{shipp18,ibata19a,grillmair95}, the other 
detections (in NGC2298, NGC6341, NGC6362, NGC7099) are new discoveries.  

The orientation of the tails follows the predictions of dynamical models: in most of 
the GCs of the sample, the inner density contours are preferentially aligned with the 
direction of the Galactic center (i.e. the direction of the Galactic potential 
gradient) which determines the elongation of the Roche lobe. On the other hand, at large 
distances the GCs with the most extended features show tidal tails aligned with the 
orbital path of the cluster. In spite of the uncertainties in the Galactic potential, 
GC distances and velocities, this last evidence is statistically significant when the whole set of GCs is considered.
In principle, the morphology of the tidal tails could be used to constrain the 
shape of the Galactic potential (in particular its flattening and the halo-to-disk 
normalisation). However, for this task a larger sample of GCs with tidal tails 
extending over a wider area than that analysed here would be needed.

For the same set of 5 GCs, the fraction of stars contained in the tidal tails has 
been used to make the first observational estimate of the present-day mass-loss rate.
Till now, these rates have been only predicted simulating the long-term evolution 
of satellites within a Galactic potential through numerical simulations or 
analytical prescriptions \citep{gnedin97,dinescu99,allen06}. 
Instead, in this work the destruction rate is directly linked to the observed 
fraction of stars contained in the visible portion of the tails. Although this 
estimate is subject to the uncertainties in the amplitude of the low-surface 
brightness tails and in the normalisation of the cluster profile, and 
neglects many dynamical processes \citep[like e.g. mass-segregation;][]{balbinot18}, it depends 
only weakly on the modelled orbit and it is linked to an observational feature.
Although the small sample size does not allow to draw any firm 
conclusion, the correlation of the destruction rate with the half-mass relaxation 
time is stronger than that with the tidal shocks strength, suggesting a dominant role 
of internal over external dynamical evolution in these GCs. This scenario is in 
agreement with the evidence of a tight anticorrelation between the mass function 
slope and half-mass relaxation time observed in a sample of 29 GCs by \citet{sollima17}, 
since dynamically evolved GCs would experience a strong 
mass loss which efficiently deplete the low-mass end of their 
mass-functions \citep{baumgardt03}
while producing at the same time prominent tidal tails features. 
However, further studies based on larger statistical samples 
are needed to clarify the significance of the above correlations.

The exceptional efficiency of the algorithm presented here is mainly due to the 
increase of the dimensionality of the available parameter space provided by Gaia.
Still, the limiting factor is the inhomogeneous sampling at faint magnitudes of this survey
which forces the adoption of a cut at relatively bright magnitudes. 
The next Gaia data releases will increase the power of tidal features detection because 
of the foreseen reduced uncertainties and better spatial sampling. 

\section*{Acknowledgments}

I warmly thank Michele Bellazzini for 
useful discussions and Holger Baumgardt for providing his unpublished destruction rates.
I also thank the anonymous referee for his/her helpful comments and suggestions that 
improved my paper.

\label{lastpage}
\end{document}